\pdfoutput=1

\documentclass[twocolumn, times]{aastex63} 
\usepackage{natbib}
\usepackage{amsmath}
\usepackage{amssymb}
\usepackage{subfigure}
\usepackage{url}
\usepackage{longtable}

\renewcommand{\vec}[1]{ {\mathbf #1} }

\newcommand{\grad}{ {\bf \nabla } }

\newcommand{\Eq}{{Equation}}
\newcommand{\Eqs}{{Equations}}
\newcommand{\Fig}{{Figure}}

\newcommand{\dive}{\nabla\cdot}

\newcommand{\gradxy}{\nabla_{\perp}}
\graphicspath{figs/}

\shorttitle{Simulation of AR~12158}
\shortauthors{Jiang et al.}

\begin{document}

\title{Data-driven MHD simulation of a sunspot rotating active region
  leading to solar eruption}

\author[0000-0002-0786-7307]{Chaowei Jiang} \affiliation{Shenzhen Key
  Laboratory of Numerical Prediction for Space Storm, Institute of
  Space Science and Applied Technology, Harbin Institute of
  Technology, Shenzhen 518055, China} \affiliation{Key Laboratory of
  Solar Activity and Space Weather, National Space Science Center,
  Chinese Academy of Sciences, Beijing 100190, China}

\author{Xueshang Feng} \affiliation{Shenzhen Key Laboratory of
  Numerical Prediction for Space Storm, Institute of Space Science and
  Applied Technology, Harbin Institute of Technology, Shenzhen 518055,
  China} \affiliation{Key Laboratory of Solar Activity and Space
  Weather, National Space Science Center, Chinese Academy of Sciences,
  Beijing 100190, China}

\author{Xinkai Bian} \affiliation{Shenzhen Key Laboratory of Numerical
  Prediction for Space Storm, Institute of Space Science and Applied
  Technology, Harbin Institute of Technology, Shenzhen 518055, China}

\author{Peng Zou} \affiliation{Shenzhen Key Laboratory of Numerical
  Prediction for Space Storm, Institute of Space Science and Applied
  Technology, Harbin Institute of Technology, Shenzhen 518055, China}

\author{Aiying Duan} \affiliation{School of Atmospheric Sciences, Sun
  Yat-sen University, Zhuhai 519000, China}

\author{Xiaoli Yan} \affiliation{Yunnan Observatories, Chinese Academy
  of Sciences, Kunming 650216, China}

\author{Qiang Hu} \affiliation{Center for
  Space Plasma and Aeronomic Research, The University of Alabama in
  Huntsville, Huntsville, AL 35899, USA}

\author{Wen He}\affiliation{Center for Space Plasma and Aeronomic
  Research, The University of Alabama in Huntsville, Huntsville, AL
  35899, USA}

\author{Xinyi Wang} \affiliation{Key Laboratory of Solar Activity and
  Space Weather, National Space Science Center, Chinese Academy of
  Sciences, Beijing 100190, China}

\author{Pingbing Zuo} \affiliation{Shenzhen Key Laboratory of
  Numerical Prediction for Space Storm, Institute of Space Science and
  Applied Technology, Harbin Institute of Technology, Shenzhen 518055,
  China} \affiliation{Key Laboratory of Solar Activity and Space
  Weather, National Space Science Center, Chinese Academy of Sciences,
  Beijing 100190, China}

\author{Yi Wang} \affiliation{Shenzhen Key Laboratory of Numerical
  Prediction for Space Storm, Institute of Space Science and Applied
  Technology, Harbin Institute of Technology, Shenzhen 518055, China}
\affiliation{Key Laboratory of Solar Activity and Space Weather,
  National Space Science Center, Chinese Academy of Sciences, Beijing
  100190, China}

\begin{abstract}
  Solar eruptions are the leading driver of space weather,
  and it is vital for space weather forecast to understand in what
  conditions the solar eruptions can be produced and how they are
  initiated. The rotation of sunspots around their umbral center has
  long been considered as an important condition in causing solar
  eruptions. To unveil the underlying mechanisms, here we carried out
  a data-driven magnetohydrodynamics simulation for the event of a
  large sunspot with rotation for days in solar active region NOAA
  12158 leading to a major eruption. The photospheric velocity as
  recovered from the time sequence of vector magnetograms are inputted
  directly at the bottom boundary of the numerical model as the
  driving flow. Our simulation successfully follows the long-term
  quasi-static evolution of the active region until the fast eruption,
  with magnetic field structure consistent with the observed coronal
  emission and onset time of simulated eruption matches rather well
  with the observations. Analysis of the process suggests that through
  the successive rotation of the sunspot the coronal magnetic field is
  sheared with a vertical current sheet created progressively, and
  once fast reconnection sets in at the current sheet, the eruption is
  instantly triggered, with a highly twisted flux rope originating
  from the eruption. This data-driven simulation stresses magnetic
  reconnection as the key mechanism in sunspot rotation leading to
  eruption.
\end{abstract}


\section{Introduction}
\label{sec:intro}

It is commonly believed that solar eruptions, such as solar flares and
coronal mass ejections (CMEs), are explosive disruption and rapid
energy release of the magnetic field configurations in the solar
corona. Such events could severely influence the solar-terrestrial
space environment and cause disastrous space weather. Therefore it is
vital for space weather forecast to understand in what conditions the
solar eruptions can be produced and what mechanisms can initiate the
solar eruptions. Observations show that major solar flares and CMEs
often originate in solar active regions (ARs) which harbour sunspots
where the magnetic fields are as strong as a few thousands of
Gauss. Moreover, a large number of eruption-productive ARs have been
reported with significant sunspot rotations
\citep{brownObservationsRotatingSunspots2003,
  yanRelationshipRotatingSunspots2008}, which represents an important
condition among others for producing eruptions, for example, in the
extensively studied ARs including National Oceanic and Atmospheric
Administration (NOAA) 10930
\citep{minRotatingSunspotAR2009,zhangInteractionFastRotating2007},
11158
\citep{jiangRAPIDSUNSPOTROTATION2012,vemareddyROLEROTATINGSUNSPOTS2012},
11429 \citep{zhengSunspotsRotationMagnetic2017}, 12158
\citep{biObservationReversalRotation2016,
  vemareddySUNSPOTROTATIONDRIVER2016}, and 12673
\citep{yanSimultaneousObservationFlux2018,yanSuccessiveXclassFlares2018},
etc. Observations show that the angular
rotational speed for many sunspots is on average a few degrees per
hour and the total rotation degree is mostly between 40--200$^{\circ}$
over periods of 3--5
days~\citep{brownObservationsRotatingSunspots2003}.
Such rotational motion of sunspots has long been considered as an
important process in association with generation of solar eruptions,
because it is an efficient mechanism for transporting free magnetic
energy and helicity from below the photosphere into the
corona~\citep{stenfloMechanismBuildupFlare1969,
  barnesForceFreeMagneticFieldStructures1972}. For example, by modelling of a solar
flare from 13 May 2005, it has been shown that the sunspot rotation of
the source AR dominates the energy accumulation for the flare event
\citep{kazachenkoSUNSPOTROTATIONFLARE2009}. In fact, such sunspot
rotation alone can store sufficient energy to power a very large
flare.

Since the coronal magnetic field is difficult to observe and measure
directly, numerical simulations in the framework of
magnetohydrodynamics have been used to study how sunspot rotating ARs
cause eruptions. Early numerical experiments starting with a
symmetrical bipolar arcade show that, by rotating the footpoints of
the bipole (which is analogous to a pair of sunspots), the continuous
twisting of the core field can lead to a strong expansion (or ``fast
opening'') of the field~\citep{amariVeryFastOpening1996,
  torokEvolutionTwistingCoronal2003,
  aulanierEquilibriumObservationalProperties2005}. However, such
expansion cannot be taken as solar eruption since there is no
impulsive release (increase) of magnetic (kinetic) energy, and the
twisted flux tube can always relax smoothly to an equilibrium if the
driving rotational velocities are turned off, therefore without
evidence for instability or a loss of equilibrium, as is carefully
analyzed by~\citet{aulanierEquilibriumObservationalProperties2005}.
\citet{torokInitiationCoronalMass2013} designed an interesting
simulation with successful eruption by first place a magnetic flux
rope enveloped within an arcade and then apply rotation to the
footpoints of the arcade. The configuration is initially in
equilibrium, and the sunspot rotation can cause the envelope arcade
overlying the flux rope to inflate. This weakens the confining effect
on the flux rope, allowing it to ascend slowly until an eruption is
initiated by torus instability of the flux rope. This mechanism may
explain a filament eruption caused by sunspot rotation in AR~10898, as
suggested by the data-inspired simulation also carried out
in~\citep{torokInitiationCoronalMass2013}. Note that in this
simulation, it does not show how the flux rope forms, and the rotating
sunspot does not energize the key structure of eruption (i.e., the
flux rope). A data-inspired simulation is also performed
by~\citet{jingUnderstandingInitiationM22021} for the eruption in
AR~12665 by first reconstructing a potential field and then applying
rotational motion of sunspot at the photosphere along with flux
emergence at the polarity inversion line (PIL). As a consequence, a sigmoidal structure formed
with an overlying MFR created and rose to erupt like a CME.  However,
their simulation may not be consistent with the commonly-accept
magnetic energy storage-and-release scenario for solar eruptions,
since the magnetic free energy show a continuous increase with nearly
the same rate throughout the whole simulation time. Also, the kinetic
energy does not exhibit a transition from slow pre-eruption evolution
to fast eruption but keeps continuous increasing with nearly the same
rate.  Thus, the role of the sunspot rotation in causing the eruption
remains unclear in their simulation.  In a recent work, we have also
performed a data-inspired MHD simulation to study the eruption
mechanism in the
AR~10930~\citep{wangMHDSimulationHomologous2022a}. The simulation is
also started with a potential magnetic field reconstructed from the
observed magnetogram and then rotational motion is applied to the
positive sunspot of the AR to mimic the observed rotation. That
simulation successfully showed that the sunspot rotation produced
homologous eruptions having reasonable consistency with observations
in relative strength, energy release, spatial features (such as
pre-eruption sigmoid and flare ribbons), and time intervals of
eruptions. In addition, the rotation angle of the sunspot before the
eruption in the simulation is also close to the observed value. The
simulation shows that as driven by the sunspot rotation, current sheet
is formed above the main PIL between the two major magnetic polarities
of the AR, and the eruptions are triggered by fast reconnection in the
pre-eruption formed current sheet.

Up to the present, there is still no data-driven MHD
simulation~\citep{jiangDatadrivenModelingSolar2022} of solar eruption
in which the rotational flow is directly derived from observations,
since in all the aforementioned simulation studies the rotation flow
is specified in an ad-hoc way. In this paper, we carried out a
data-driven MHD simulation for solar AR NOAA 12158 which contains a
prominent rotating sunspot that eventually produced a X-class eruptive
flare. Importantly, the photospheric velocity as recovered from the
time sequence of vector magnetograms are inputted directly as the
driving flow. Our simulation follows the long-term evolution of the AR
until the eruption, and onset time of eruption the simulation matches
rather well with the observations. The simulation shows that, before
the eruption, the nonpotentiality of the coronal magnetic field, as
measured by the ratio of the total magnetic energy to the
corresponding potential field energy, is increased monotonically by
the surface rotation flow, while the kinetic energy keeps a small
value, as the MHD system evolves quasi-statically. At a critical time,
there is a clear transition from the quasi-static state to an eruptive
phase in which the kinetic energy impulsively rises and the magnetic
energy releases quickly. Such a key transition is associated with a
vertical current sheet created progressively in the sheared arcade as
driven by the sunspot rotation. The eruption is triggered by fast
reconnection in the current sheet, and a highly twisted flux rope
originates from the eruption, forming a CME. In the following, we will
first give an observational analysis of the studied event in
Section~\ref{obs}, then describe our numerical modeling settings in
Section~\ref{model} and show the simulation results in
Section~\ref{res}, and finally conclude and give discussions in
Section~\ref{con}.

\begin{figure*}[htbp]
  \centering \includegraphics[width=0.8\textwidth]{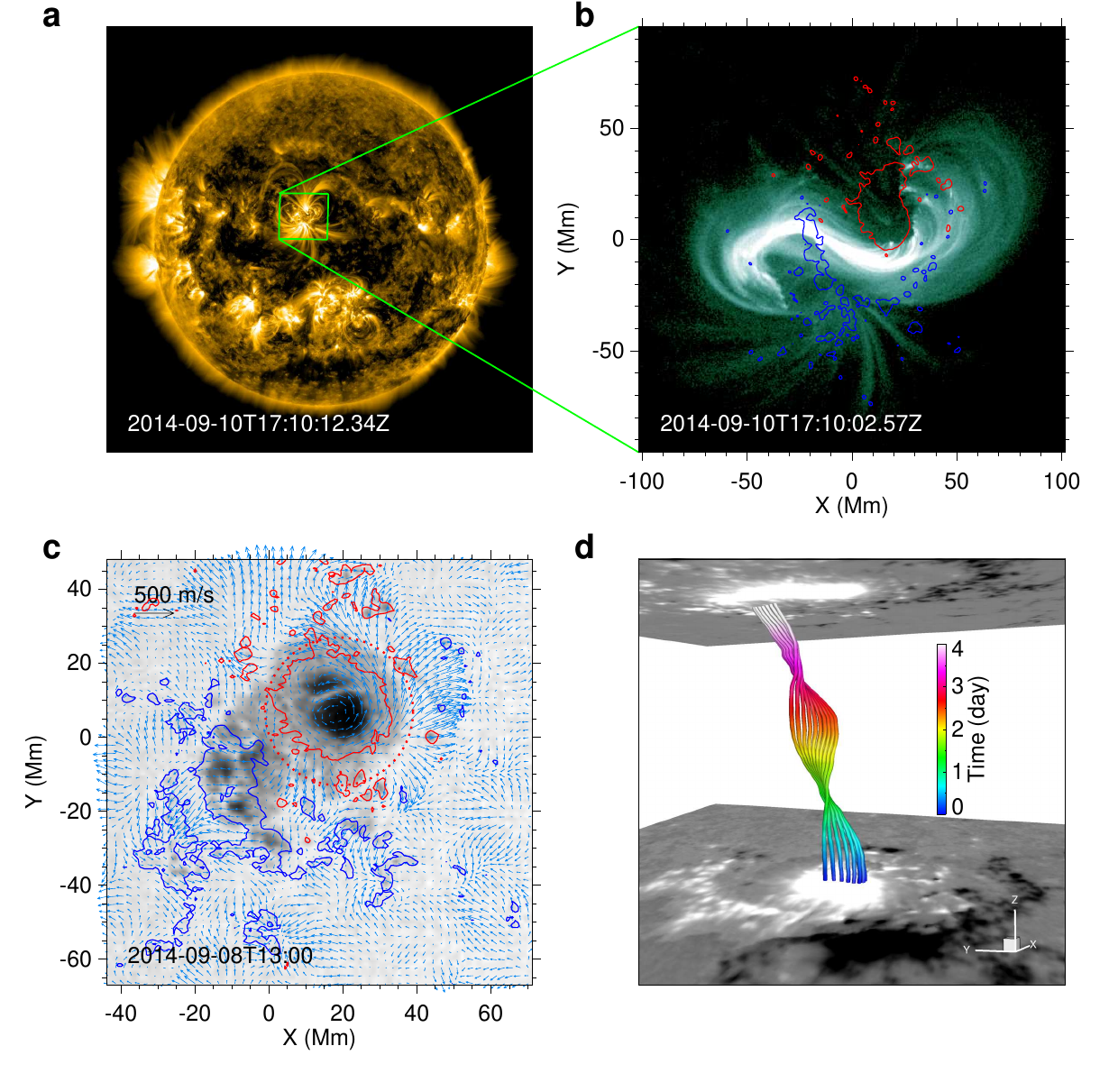}
  \caption{\textbf{Observation of coronal structure and the rotational
      sunspot in AR~12158.} \textbf{a}, Full-disk image of the Sun
    taken by SDO/AIA in 171~{\AA} channel at 10~min before onset of
    the X1.6 flare. The boxed region denotes the location of the
    target AR. \textbf{b}, The enlarged view of the AR in SDO/AIA
    94~{\AA} channel which presents a coronal sigmoid with reverse S
    shape. The colored curves are shown for the contour lines of the
    vertical magnetic field $B_z$, with red (blue) representing
    $B_z = 500$ $(-500)$~G. \textbf{c}, Surface velocity field (the
    vectors colored in green) overlaid on the SDO/HMI continuum image
    of the AR. The main sunspot is denoted by the circle with a radius
    of $20$~Mm. \textbf{d}, 3D shape of the trajectories of the
    surface rotational flow with the $z$ axis representing time
    direction, and the color also denotes time. The bottom and top
    surfaces represent the magnetogram observed at 00:00~UT on 7 and
    00:00~UT on 11 of September 2014, respectively.}
   \label{fig_rotation}
\end{figure*}

\section{Overview of the event}\label{obs}

AR NOAA~12158 first appeared on the solar disk on 5 September 2014,
and is already in its decaying phase with continual decrease of the
total unsigned magnetic flux. The AR has an overall bipolar magnetic
configuration, and it has a leading sunspot with continuous
counterclockwise rotation for days of 6--11 September 2014 during its
passage on the solar disk (\Fig~\ref{fig_rotation} and Supplementary
Video 1). \Fig~\ref{fig_rotation}d shows the trajectories of the
surface flow (as derived from time sequence of vector magnetograms,
see Section~\ref{vel}) which clearly indicate the rotation of the
sunspot. The rotation degree of the sunspot has also been estimated
in~\citet{brownSemiAutomaticMethodMeasure2021} and
\citet{vemareddySUNSPOTROTATIONDRIVER2016}, which show that it has
rotated by around $200^{\circ}$ in the days. As dominated by the
sunspot rotation, the coronal configuration of the AR is driven to
form an inverse S-shaped structure with hot emission as seen in the
94~{\AA} image (\Fig~\ref{fig_rotation}b) of Atmospheric Imaging
Assembly (AIA) onboard Solar Dynamics Observatory (SDO), i.e., a
sigmoid, surrounded by cooler large-scale loops as seen in AIA
171~{\AA} (\Fig~\ref{fig_rotation}a). At around 17:00~UT on 10
September, the AR produced a
Geostationary-Operational-Environmental-Satellite (GOES)
X1.6 flare, which
was accompanied with a global eruption of the AR resulting in a halo
CME~\citep{vemareddySUNSPOTROTATIONDRIVER2016}. This flare has
attracted a lot of attentions in previous studies due to the complex
evolution of its flare ribbons and
loops~\citep{liQUASIPERIODICSLIPPINGMAGNETIC2015,
  dudikSLIPPINGMAGNETICRECONNECTION2016,gouCompleteReplacementMagnetic2023}.
Our simulation is aimed to follow the coronal magnetic evolution from
00:00~UT on 8 September 2014 until this major eruption.


\section{Photospheric velocity recovering}\label{vel}
Our simulation is driven by the surface velocity at the photosphere,
which can be derived from the time series data of vector
magnetograms. Here we used such a velocity-recovering method called
DAVE4VM~\citep{schuckTrackingVectorMagnetograms2008}. It is a
differential affine velocity estimator (DAVE) designed for vector
magnetogram (VM), which uses a variational principle to minimize
statistically deviations in the magnitude of the magnetic induction
equation. The vector magnetograms are provide by the SDO/HMI SHARP
data with cadence of 12 minutes and pixel size of 1~arcsec (by
rebinning the original data with pixel size of 0.5~arcsec). Since
there are a few time gaps (specifically, 4 time gaps of around 2 hours
which are, respectively, 06:12~UT to 08:48~UT on 8 September, 06:00~UT
to 08:48~UT on 9 September, 00:12~UT to 03:00~UT and 06:00 UT to 08:48
UT on 10 September) in our studied time interval, we first filled the
data gap using linear interpolation in the time domain to generate a
complete time series from 00:00~UT on 8 September 2014 to 00:00~UT on
11 September 2014. Then we input the time series of vector magnetogram
into the DAVE4VM code. We set the window size of sampling, a key
parameter in the DAVE4VM code, as its optimized value of 19~pixels
\citep{liuMAGNETICENERGYHELICITY2012,
  liuTHREEDIMENSIONALMAGNETICRESTRUCTURING2014}.

After obtaining the surface velocity, we made a correction by removing
the velocity component parallel to the magnetic field, since this
field-aligned velocity is artificial and makes no contribution in the
magnetic induction equation. To reduce the data noises, the time
series of flow maps are smoothed in both the time and space domains,
with a Gaussian FWHM of 2 hours for time (i.e., 10 times of the data
cadence) and 6~arcsec for both $x$ and $y$ directions, respectively,
After this preparation, the velocity will be entered into the
data-driven model. \Fig~\ref{fig_rotation}c shows a snapshot of the
surface velocity after this smoothing. The speed of the flow is
generally a few hundreds of meters per second and the main feature is
a clear and persistent rotation of the main sunspot. Note that during
the three days the basic configuration of the photospheric magnetic
flux distribution is rather similar with only gradual dispersion as
small magnetic flux fragments, known as moving magnetic features
\citep{harveyObservationsMovingMagnetic1973}, move outward from the
sunspots as advected by the moat flow (i.e., the diverging flow
existing persistently in the periphery of the sunspot).

\section{Numerical model}\label{model}
To simulate the quasi-static slow evolution of the AR until its fast
eruption, we selected 00:00~UT on 8 September 2014, a time over 65
hours before onset of the X1.6 flare, as a starting point. We first
constructed an MHD equilibrium based on a single vector magnetogram
taken by the SDO/HMI for the starting time, using an MHD-relaxation
technique. The MHD equilibrium represents a snapshot of coronal
evolution at that time. Then, with this well-established equilibrium
as the initial condition, we carried out MHD simulation as driven at
the bottom boundary by the surface velocity as prepared in
Section~\ref{vel}.


\subsection{Model equations}
The simulation is carried out by solving numerically the MHD equations
using an advanced conservation element and solution element (CESE)
method~\citep{fengNovelNumericalImplementation2007,
  jiangAMRSimulationsMagnetohydrodynamic2010b}. The control equations
are as follows,
\begin{eqnarray}
  \label{eq:MHD}
  \frac{\partial \rho}{\partial t}+\dive (\rho\vec v) =
  -\nu_{\rho}(\rho-\rho_0),\nonumber \\
  \rho\frac{d\mathbf{v}}{d t} = -\grad p+\vec J\times \vec B+\rho\vec
  g + \nabla\cdot(\nu\rho\nabla\mathbf{v}),\nonumber\\
  \frac{\partial \vec B}{\partial t} =
  \grad \times (\vec v \times \vec B), \nonumber\\
  \frac{\partial T}{\partial t}+\nabla\cdot (T\vec v) =
  (2-\gamma)T\nabla\cdot\vec v.
\end{eqnarray}
where the electric current density $\vec J = \nabla \times \vec B$,
$\nu$ is the kinetic viscosity, and $\gamma$ is the adiabatic
index. Note that the equations are written in non-dimensionalized form
with all variables normalized by their typical values at the base of
the corona, which are, respectively, density
$\rho_s = 2.29\times 10^{-15}$~g~cm$^{-3}$, temperature
$T_s = 10^6$~K, velocity $v_s = 110$~km~s$^{-1}$, magnetic field
$B_s = 1.86$~G, length $L_s = 11.52$~Mm, and time $t_s = 105$~s.

In this simulation, an artificial source term
$-\nu_{\rho}(\rho-\rho_0)$ has been added to the continuity equation
(i.e., the first equation in \Eqs~(\ref{eq:MHD}), where $\rho_0$ is
the density at the initial time $t=0$, and $\nu_{\rho}$ is a
prescribed coefficient given as $ \nu_{\rho} = 0.05 v_{\rm A} $
($v_{\rm A} = B/\sqrt{\rho}$ is the Alfv{\'e}n speed). This term is
used to avoid ever decreasing of the density in the strong magnetic
field region, an issue often encountered in simulations handling very
large magnetic field gradients and at the same time with very low
plasma $\beta$~\citep{jiangMHDModelingSolar2021}. It can maintain the
maximum Alfv{\'e}n speed in a reasonable level, which may otherwise
increase and make the iteration time step smaller and smaller and the
long-term simulation unmanageable. This source term is actually a
Newton relaxation of the density to its initial value by a time scale
of
\begin{equation}
 \tau_{\rho} = \frac{1}{\nu_{\rho}} = 20 \tau_{\rm A},
\end{equation}
where $\tau_{\rm A} = 1/v_{\rm A}$ is the Alfv{\'e}n time with length
of $1$ (the length unit $L_s$). Thus it is sufficiently large to avoid
influence on the fast dynamics of Alfv{\'e}nic time scales. We have
run test simulations with much larger values of $\tau_{\rho}$, which
give almost the same evolution in both kinetic and magnetic energies,
but the time step decreases substantially and therefore the whole
simulation will demand a much longer computational time. Using a
diffusion term of the density to smooth its profile has also been
considered in other simulations for the same
purpose~\citep{aulanierFORMATIONTORUSUNSTABLEFLUX2010,aulanierEquilibriumObservationalProperties2005}.

Similar to our previous
works~\citep{jiangFundamentalMechanismSolar2021,
  jiangMHDModelingSolar2021}, we chose to not use explicit resistivity
in the magnetic induction equation, but magnetic reconnection can
still be triggered through numerical diffusion when a current layer is
sufficiently narrow with thickness close to the grid resolution. By
this, we achieved an effective resistivity as small as we can with a
given grid resolution, and also mimicked the current-density-dependent
resistivity as required for fast Petscheck-type reconnection. For
simplicity, the adiabatic index is set as $\gamma=1$ in the
{temperature equation}, which thus reduces to an isothermal
process. Although in this case {the temperature equation could be
  simply discarded} by setting the temperature as a constant, we still
kept the full set of equations in our code {with solving the
  temperature equation}, which can thus describes either the
isothermal or adiabatic process {depending on} particular values of
$\gamma$. The kinetic viscosity $\nu$ is given with different values
when needed, which is described in the following sections.

\subsection{Grid setting and numerical boundary conditions}
For the purpose of minimizing the influences introduced by the side
and top numerical boundaries of the computational volume, we used a
sufficiently large box of
$(-32, -32, 0)L_{s} < (x, y, z) < (32, 32, 64)L_{s}$ embedding the
field of view of the magnetogram of
$(-8.75, -8.25)L_{s} < (x, y) < (8.75, 8.25)L_{s}$, and the simulation
runs are stopped before the disturbance by the eruption reaches any of
the side and top boundaries. The full computational volume is resolve
by a non-uniform block-structured grid with adaptive mesh refinement
(AMR), in which the highest and lowest resolution are
$\Delta x = \Delta y = \Delta z = 1/16L_{s}$ (corresponding to
$1$~arcsec or 720~km, matching the resolution of the vector
magnetogram) and $1/2L_{s}~$, respectively. The AMR is controlled to
resolve with the smallest grids the regions of strong magnetic
gradients and current density, particularly near the current
sheet. The magnetic field outside of the area of the magnetograms on
the lower boundary is given as zero for the vertical component and
simply fixed as the potential field for the transverse components. On
the side and top boundaries, since the simulation runs are stopped
before the disturbance by the eruption reaches any of these boundaries
on which all the variables do not evolve, we thus fixed the plasma
density, temperature, and velocity as being their initial values. But
to avoid numerical errors of magnetic divergence accumulated on these
boundaries, the tangential components of magnetic field are linearly
extrapolated from the inner points, while the normal component is
modified according to the divergence-free condition.

\begin{figure*}[htbp]
  \centering \includegraphics[width=\textwidth]{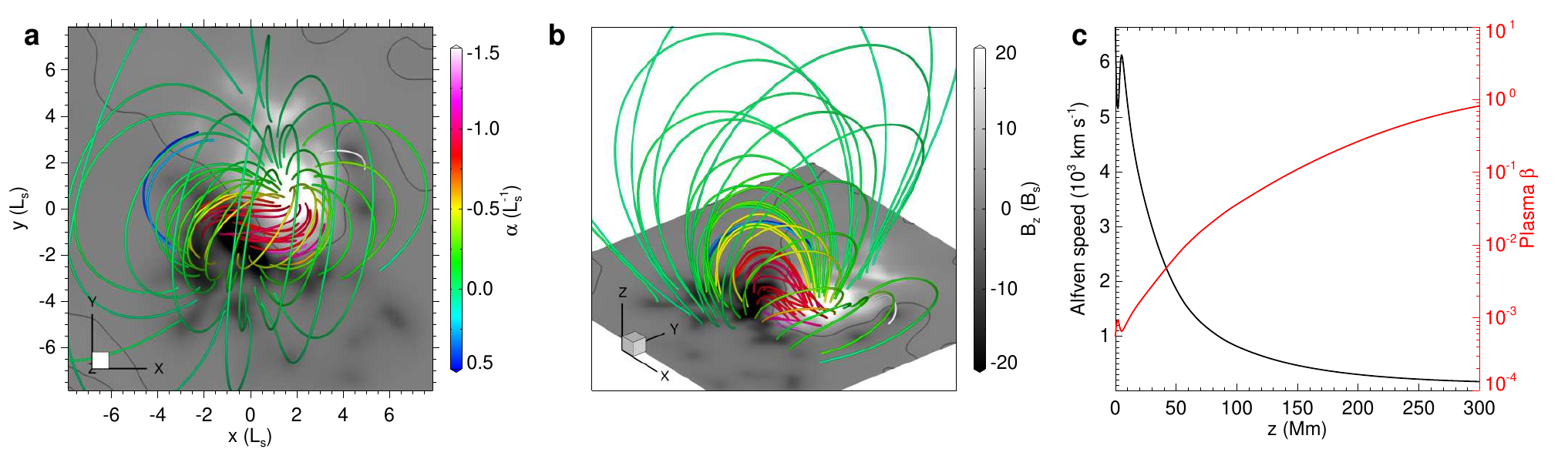}
  \caption{\textbf{The initial state of the simulation.} \textbf{a},
    Magnetic field lines as seen from above. The field lines are
    false-colored by the values of the force-free factor
    $\alpha = \vec J\cdot \vec B/B^2$. The background is shown by the
    magnetic flux distribution on the bottom surface, and the thick
    gray curves are shown for the PIL, i.e., where $B_z =
    0$. \textbf{b}, Same as \textbf{a} but seen in a 3D perspective
    view. \textbf{c}, Profiles of Alfv{\'e}n speed and plasma $\beta$
    with height along a vertical line with footpoint at the center of
    the main sunspot.}
  \label{initial_state}
\end{figure*}

\subsection{Construction of an initial MHD equilibrium}
\label{initial}
To initialize the surface flow-driven simulations, an MHD equilibrium
is constructed based on the SDO/HMI vector magnetogram taken for time
of 00:00~UT on 8 September 2014. Such an equilibrium is crucial for
starting our subsequent surface flow-driven evolution, since the
subsequent evolution might be influenced greatly by the unbalanced
force in the initial field if it is not in equilibrium. Beforehand we
preprocessed the vector magnetogram use a method developed
by~\citet{jiangPreprocessingPhotosphericVector2014a} and further
smoothed all the three components of the magnetic field using Gaussian
smoothing with FWHM of 6~arcsec. This is done for two reasons: on the
one hand, the preprocessing minimizes the photospheric Lorentz force
contained in the vector magnetogram, which is helpful for reaching a
more force-free equilibrium
state~\citep{wiegelmannPreprocessingVectorMagnetograph2006}; on the
other hand, the smoothing effectively filters out the small-scale
magnetic structures that cannot be sufficiently resolved in our
simulation, and it also mimics the effect of magnetic field expansion
from the photosphere to the base of the
corona~\citep{yamamotoPREPROCESSINGMAGNETICFIELDS2012}, since the
lower boundary of our simulation is assumed to be the coronal base
rather than directly the
photosphere~\citep{jiangTestingDatadrivenActive2020,
  jiangDatadrivenMagnetohydrodynamicModelling2016}.

We constructed the MHD equilibrium based on an MHD-relaxation approach
consisting of two steps~\citep{jiangMHDModelingSolar2021}. In the
first step, a potential magnetic field $\vec B_{\rm pot}$ extrapolated
from the vertical component (i.e., $\mathcal{B}_z$) of the
preprocessed and smoothed vector magnetogram
$(\mathcal{B}_x, \mathcal{B}_y, \mathcal{B}_z)$, along with an initial
plasma as the background atmosphere were input into the MHD model. For
the initial plasma, we used an isothermal gas in hydrostatic
equilibrium. It is stratified by solar gravity with a density
$\rho=\rho_{s}$ at the bottom and a uniform temperature of
$T=T_{s}$. With the plasma configured by typical coronal density and
temperature, we chose to reduce the original magnetic field strength
by a factor of $20$, such that the maximum of magnetic field strength
in normalized value is approximately $50\sim 100$ in the model during
the evolution. If using the original values of magnetic field, its
strength (and the characteristic Alfv{\'e}n speed) near the lower
surface is too large, and will put a too heavy burden on computation
since the time step of our simulation is limited by the CFL condition.

With these initial conditions, we modified the transverse magnetic
fields on the bottom boundary incrementally in time (using linear
extrapolation with a duration of $t=t_{s}$) from that of the potential
field $\vec B_{\rm pot}$ to that of the vector magetogram
$(\mathcal{B}_x, \mathcal{B}_y, \mathcal{B}_z)$. The process drives
the coronal magnetic field to evolve away from the initial potential
state, since the change of the transverse field injects electric
currents and thus Lorentz forces, which induce motions in the
computational volume. In this phase all other variables on the bottom
boundary are simply fixed, thus the plasma remains to be motionless
there. Although this procedure is somewhat un-physical since the
Lorentz force will also introduce nonzero flows on the bottom
boundary, it provides a simple and `safe' way (avoiding numerical
instability) to bring the transverse magnetic field into the MHD
model. Once the magnetic field on the bottom surface is identical to
that of $(\mathcal{B}_x, \mathcal{B}_y, \mathcal{B}_z)$, the MHD
system is then allowed to relax to an equilibrium with all the
variables (including the magnetic field) on the bottom boundary
fixed. {We note that this process is a classic way of reconstructing
  nonlinear force-free coronal field by using MHD-relaxation
  approach~\citep[e.g.,][]{valoriMagnetofrictionalExtrapolationsLow2007,
    jiangEXTRAPOLATIONSOLARCORONAL2013a,
    guoMAGNETOFRICTIONALMODELINGCORONAL2016,
    inoueMAGNETOHYDRODYNAMICSIMULATIONX22014}.  With this step, the
  field reaches an approximately force-free state with all three
  components of magnetic field at the bottom surface matching the
  magnetogram.}  To avoid a too large velocity in this phase such that
the system can relax fast, we set a relatively large kinetic viscosity
coefficient, which is given by $\nu = 0.5\Delta x^{2}/\Delta t$ (where
$\Delta x$ is the local grid spacing and $\Delta t$ the local time
step as determined by the CFL condition with the fastest magnetosonic
speed). Actually this is the largest viscosity one can use with a
given grid size $\Delta x$ and time step $\Delta t$, because the CFL
condition for a purely diffusive equation with diffusion coefficient
$\nu$ requires $\Delta t \le 0.5 \Delta x^{2}/\nu$. The relaxation
phase takes a time of $t=20t_{s}$ with the average relative residual
of magnetic field in the whole volume between two consecutive time
steps reduced to a sufficiently small level of below $10^{-5}$. {The
  magnetic energy as obtained in this step is $1.43 E_0$ where $E_0$
  is the corresponding potential field energy. The residual kinetic
  energy is around $0.003$ times of the magnetic energy. }

In the second step, we carried out a `deeper' relaxation by running
the model again but started with the relaxed magnetic field obtained
in the first step and the initially hydrostatic plasma. We reduced the
kinetic viscosity to $\nu = 0.05\Delta x^{2}/\Delta t$, i.e., an order
of magnitude smaller than that used in the first step, which
corresponds to a Reynolds number of $10$ for the length of a grid cell
$\Delta x$. Furthermore, the magnetic field at the bottom boundary is
allowed to evolve in a self-consistent way with assumptions that the
bottom boundary is a perfectly line-tying and fixed (i.e.,
$\vec v = 0$) surface of magnetic field lines. Note that such a
line-tying condition does not indicate that all magnetic field
components on the boundary are fixed, because even though the velocity
$\vec v$ is given as zero on the bottom boundary, it is not
necessarily zero in the neighboring inner points. To self-consistently
update the magnetic field, we solve the magnetic induction equation on
the bottom boundary. Slightly different from the one in the main
equations~(\ref{eq:MHD}), the induction equation at the bottom surface
is given by
\begin{equation}
\label{photo_B_equ}
  \frac{\partial \vec B}{\partial t} =  \grad \times (\vec v \times \vec B) +
  \eta_{\rm p} \gradxy^{2} \vec B,
\end{equation}
where we added a surface diffusion term defined by a surface Laplace
operator as
$\gradxy^{2} =\frac{\partial^{2}}{\partial
  x^{2}}+\frac{\partial^{2}}{\partial y^{2}}$
with a small resistivity for numerical stability near the PIL
$\eta_{\rm p} = 1\times 10^{-3} e^{-B_z^2}$, since the photospheric
magnetic fields often have the strongest gradient across the main
PIL. The surface induction \Eq~(\ref{photo_B_equ}) in the code is
realized by second-order difference in space and first-order forward
difference in time. Specifically, on the bottom boundary (we do not
use ghost cell\footnote{In our code, all the variables are specified
  on the grid nodes (i.e., corners of cells), and on the boundary
  surfaces no ghost cell is used. Therefore the data from the
  observations at the photosphere (i.e., magnetogram data and the
  derived flow field) are given on the bottom surface, exactly the
  $z=0$ surface.}), we first compute $\vec v\times \vec B$, and then
use central difference in horizontal direction and one-sided
difference (also 2nd order) in the vertical direction to compute the
convection term $\grad \times (\vec v \times \vec B)$. The surface
Laplace operator is also realized by central difference.

{The magnetic energy during this step changes very little by only $3$
  percent (from $1.43 E_0$ to $1.39 E_0$), while the residual kinetic
  energy is reduced by over three times, achieving a small ratio of
  kinetic energy to magnetic energy of well below
  $10^{-3}$}. \Fig~\ref{initial_state}a and b shows the 3D magnetic
field lines of the final relaxed MHD equilibrium. Note that the field
lines are pseudo-colored by the values of the force-free factor
defined as $\alpha = \vec J\cdot \vec B/B^2$, which indicates how much
the field lines are non-potential. For a force-free field, this
parameter is constant along any given field line. As can be seen, the
magnetic field is close to a force-free one since the color is nearly
the same along any single field line. In the core of the
configuration, the field lines are sheared significantly along the
PIL, thus having large values of $\alpha$ and current density. On the
other hand, the overlying field is almost current-free or
quasi-potential field with $\alpha \sim 0$, and it plays the role of
strapping field that confines the inner sheared
core. \Fig~\ref{initial_state}c shows the profile of plasma $\beta$
and Alfv{\'e}n speed with height, as an example, along a vertical line
with footpoint at the center of the main sunspot. The largest
Alfv{\'e}n speed is more than $6 \times 10^3$~km~s$^{-1}$, and the
plasma $\beta$ is mostly smaller than unity below $300$~Mm with the
smallest value of $5\times 10^{-4}$, therefore satisfying well the
essential conditions of dynamics in the corona, i.e., very large
Alfv{\'e}n speed and low plasma $\beta$.

\subsection{The velocity-driven approach}
Here to save computing time, the cadence of the input flow maps, which
is originally 12~min, was increased by $68.6$ times when inputting
into the MHD model, {and correspondingly the driving velocity is
  scaled up by $68.6$ times}. This means that a unit of time in the
simulation, $t_s$, corresponds to actually $t_s \times 68.6=7200$~s,
i.e., 2 hours, in the HMI data. Compressing of the time in HMI data is
justified by the fact that the speed of photospheric flows is often a
few of
$10^2$~m~s$^{-1}$~\citep[e.g.,][]{liuHorizontalFlowsPhotosphere2013}). So
in our model settings, the evolution speed of the boundary field, even
enhanced by a factor of $68.6$, is still smaller by two orders of
magnitude than the coronal Alfv{\'e}n speed (on the order of
$10^3$~km~s$^{-1}$), and the quick reaction of the coronal field to
the slow bottom changes should not be affected. It should be noted
that this is valid only in the quasi-static evolution process, while
in the eruption phase, the coronal field starts to evolve
significantly faster than in the quasi-static phase and the time scale
is controlled by the corona itself.

The implementation of the bottom boundary conditions is the same as
that for the second step (i.e., the `deeper' relaxation phase)
descried in Section~\ref{initial}, except that here an enhanced
surface diffusion of magnetic field is applied. Since observations
show that there is a large amount (around 30\%) of flux decrease in
the three days due to persistent flux dispersion, diffusion and
cancellation associated with the decaying of the AR, we used an
enhanced surface diffusion term to the magnetic field at the bottom
boundary, which on the one hand can simulate the overall flux
decrease, and on the other hand can avoid over steepening of magnetic
flux density at the moat of the sunspot. Otherwise in the simulation
the magnetic flux as carried outward persistently by the diverging
moat flows will pileup at the locations where the moat flows
disappear~\citep[which has been noticed
in][]{jiangMHDModelingSolar2021}. Specifically, the enhanced surface
diffusion of magnetic field is given depending on the local field
strength as
\begin{equation}\label{eq:eta_p}
  \eta_{\rm p} = 2 \times 10^{-3} + \left\{
    \begin{array} {r@{\quad;\quad}l}
      \left(\frac{B}{100}\right)^{2}\times
      10^{-2} & B \le 100\\
      10^{-2} & B \ge 100
    \end{array} \right.
\end{equation}
Such an enhanced diffusion term is used mainly to avoid a too large
magnetic flux density accumulated at the moat of the sunspot.

\begin{figure}[htbp]
  \centering
  \includegraphics[width=0.49\textwidth]{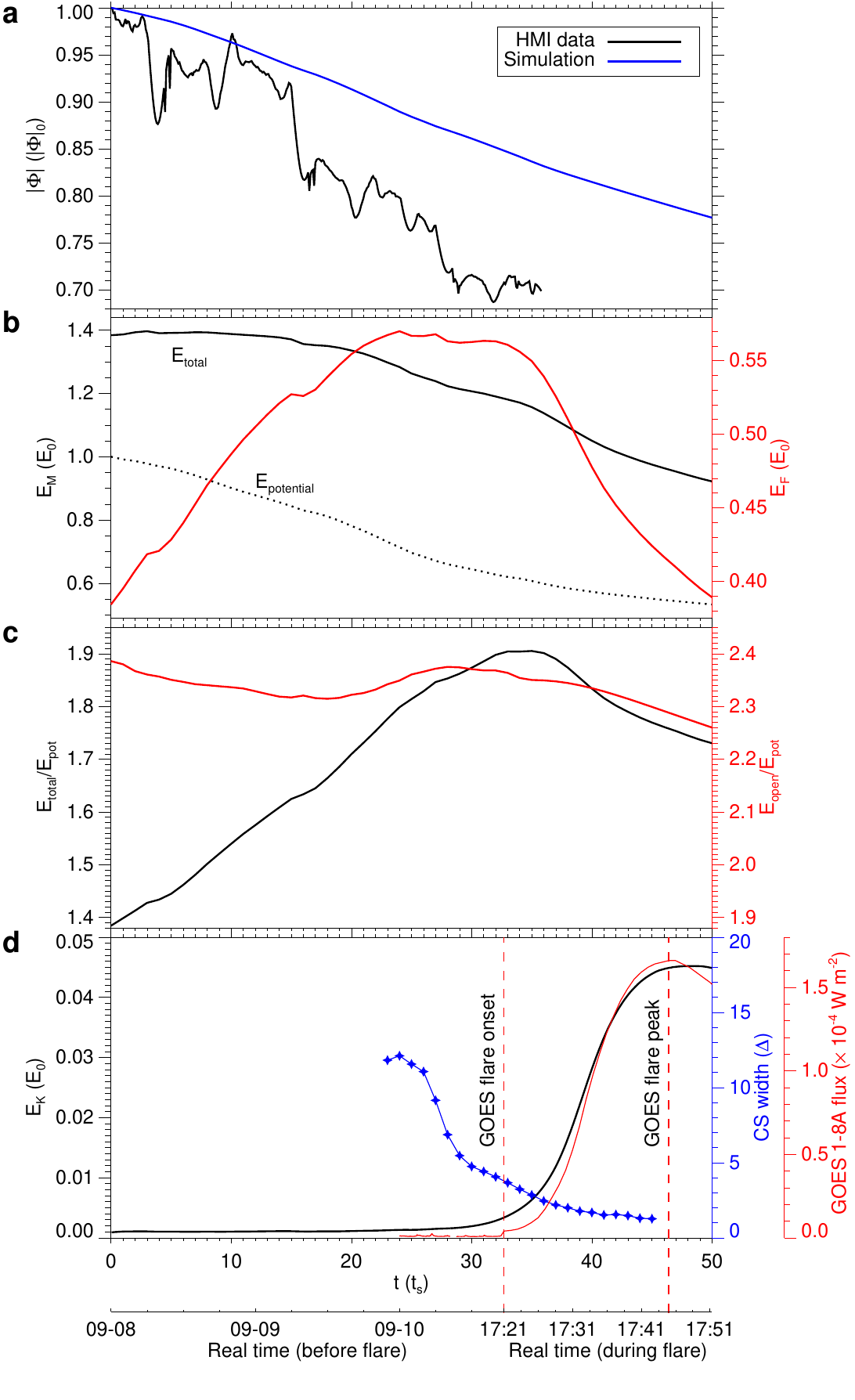}
  \caption{\textbf{Temporal evolution of different parameters in the
      data-driven V2D simulation.} \textbf{a}, Total unsigned magnetic
    flux at the bottom surface in both observation (the black curve)
    and the simulation (the blue curve). Both values are normalized by
    their initial value at $t=0$. \textbf {b}, The total magnetic
    energy is shown by the solid black line, and the potential field
    energy by the dashed black line, with $y$-axis on the left, and
    the free energy (i.e., the total magnetic energy subtracted by the
    potential field energy) is shown by the solid red line with the
    right $y$-axis. All the energies are normalized by the initial
    potential field energy at $t=0$. \textbf{c}, The degree of
    non-potentiality of the coronal field, as measured by the ratio of
    total magnetic energy $E_{\rm total }$ to the corresponding
    potential field energy $E_{\rm pot}$. For reference, the ratio of
    the open field energy to the potential field energy is shown with
    the right $y$-axis. \textbf{d}, Kinetic energies with axis in the
    left, and width of the current layer with axis in the right. The
    GOES soft X-ray flux is also shown by the red line. Note that the
    realistic time before the flare onset is scaled to the simulation
    time by a factor of $68.6$, and after the flare onset no scale is
    applied.  The two vertical dashed lines denote the flare onset
    time and peak time.}
  \label{paraevol_OBS}
\end{figure}

\begin{figure*}[htbp]
  \centering
  \includegraphics[width=\textwidth]{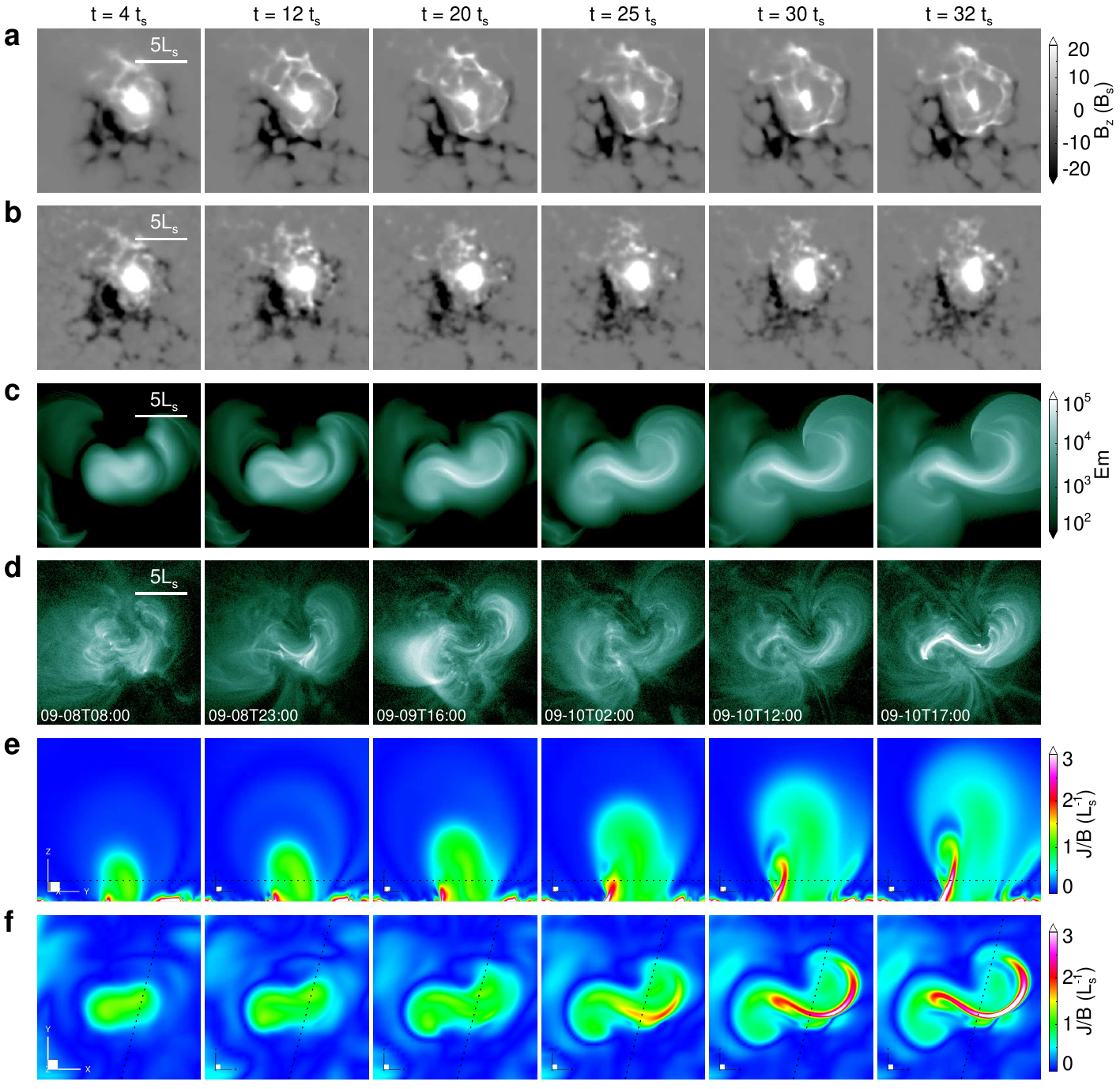}
  \caption{\textbf{Pre-eruption evolution of magnetic field and
      current density in the V2D data-driven simulation.} \textbf{a},
    The magnetic flux distribution at the bottom surface.  \textbf{b},
    {The observed magnetograms at the same times corresponding to
      those shown in a}. \textbf{c}, Synthetic images of coronal
    emission from current density. \textbf{d}, SDO/AIA 94~{\AA}
    images. \textbf{e}, Vertical cross-section of the normalized
    current density. \textbf{f}, Horizontal cross-section of the
    normalized current density. The projected location of the vertical
    cross-section in \textbf{e} is denoted by the black line in
    \textbf{f}. The height of the horizontal cross-section in
    \textbf{f} is shown by the black line in \textbf{d}.}
  \label{fig3_OBS}
\end{figure*}

\begin{figure*}[htbp]
  \centering
  \includegraphics[width=\textwidth]{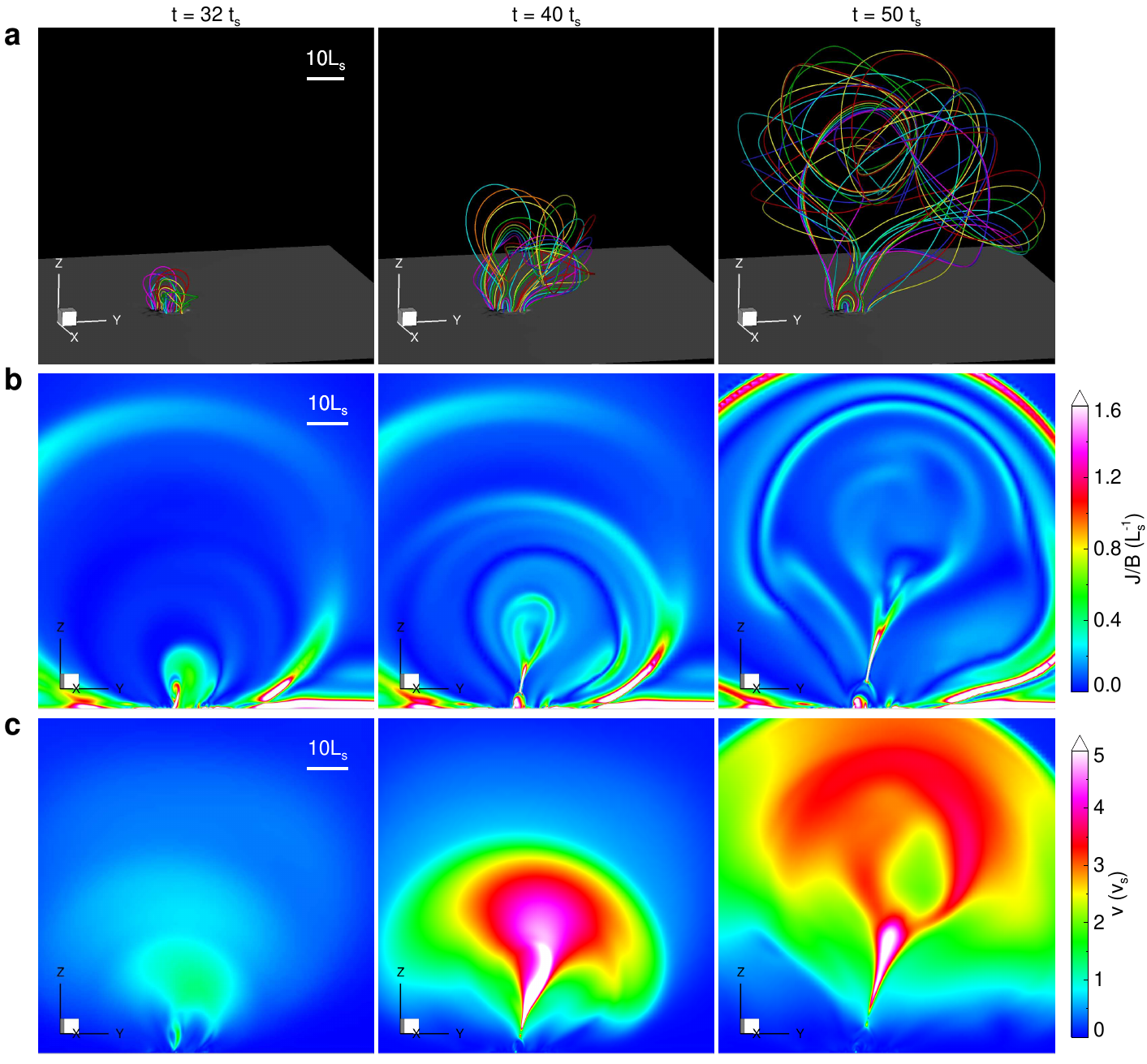}
  \caption{\textbf{Eruption in the data-driven V2D
      simulation}. \textbf{a}, Evolution of magnetic field lines,
    which are shown by the thick coloured lines, and the colours are
    used for a better visualization of the different
    lines. \textbf{b}, Vertical cross section of the normalized
    current density. \textbf{c}, Vertical cross section of the
    velocity.}
  \label{fig4_OBS}
\end{figure*}

\begin{figure*}[htbp]
  \centering
  \includegraphics[width=0.8\textwidth]{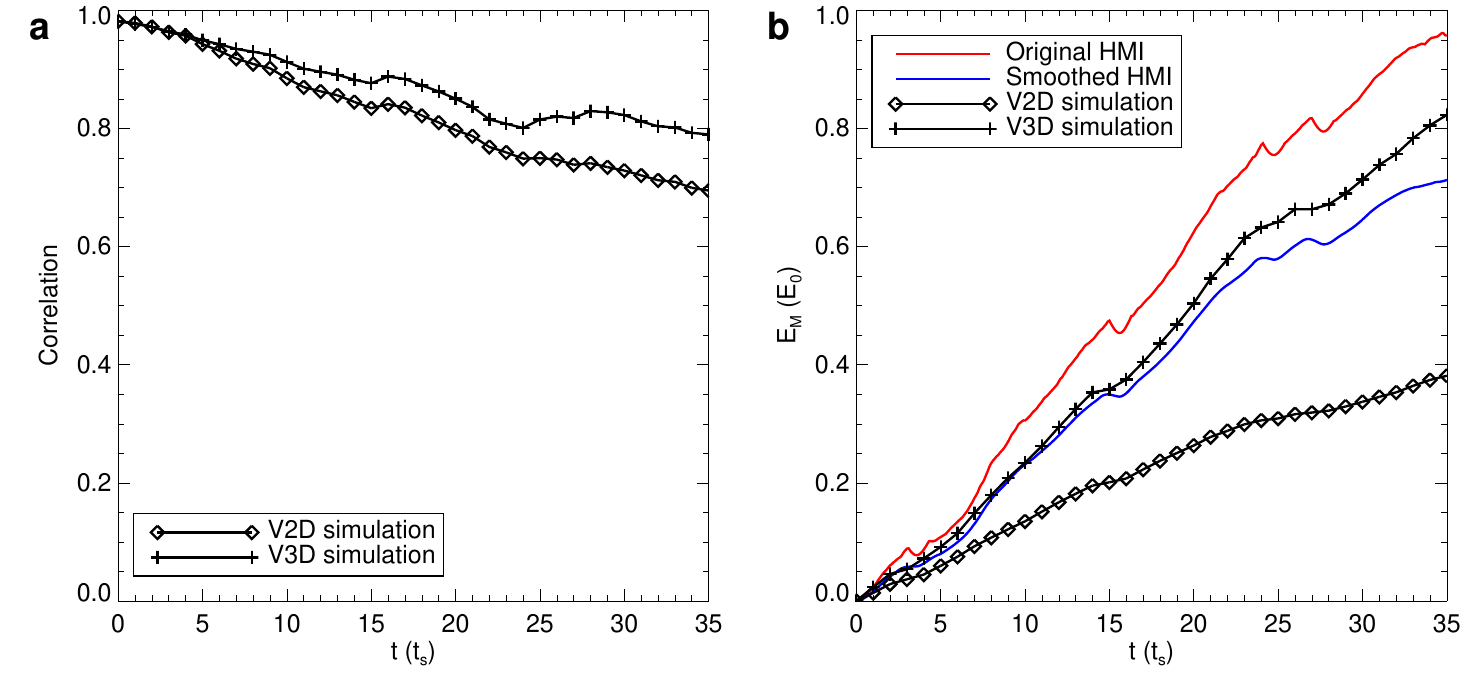}
  \caption{\textbf{Comparison of magnetic field and cumulative
      magnetic energy at the bottom boundary from the data-driven
      simulations with those from observations.}  \textbf{a}, The
    correlation coefficient (CC) between the simulated and the
    observed magnetic field $B_z$ at the bottom surface. Here the
    observed $B_ z$ is also smoothed in both time and space domains,
    with a Gaussian FWHM of 2 hours for time (i.e., 10 times of the
    data cadence) and 6~arcsec for both $x$ and $y$ directions,
    respectively.  \textbf{b}, The cumulative magnetic energy as
    computed from the bottom boundary (which is defined as
    $E = \frac{1}{\mu_0}\int_0^t \int_S \vec B \times (\vec v \times
    \vec B) \cdot d\vec s dt $ where $S$ is the boundary surface) in
    the numerical model with that derived directly from the observed
    data (using the $\vec v$ and the observed $\vec B$). The red line
    denotes the value derived from original HMI data and the blue line
    for value derived from smoothed HMI data.}
   \label{compare_mag_fig}
\end{figure*}

\section{Simulation results}\label{res}

We have carried out two data-driven simulations, one driven by only
horizontal components $(v_{x}, v_{y})$ of the photospheric velocity
field and $v_{z}=0$, referred to as V2D simulation), and the other
driven by the full 3D photospheric velocity field
($(v_{x}, v_{y}, v_{z})$, V3D simulation). The V2D simulation is used
to highlight the effect of the sunspot rotation, since the rotation is
contained in the horizontal motion, while the V3D simulation also
include the effect of flux emergence ($v_{z} > 0$) and submerge
($v_{z} < 0$) if there is.
The two data-driven simulations actually show very similar results in
evolutions of both the magnetic topology and the energies (see
Supplementary Video~2 for an overview of the V2D simulation and
Supplementary Video~3 for the V3D simulation), suggesting that the
vertical velocity is not the key factor in driving the simulation for
this AR. This is consistent with the fact that the AR is in its
decaying phase in our studied time. Therefore, below we only give an
analysis of the V2D simulation.

\Fig~\ref{paraevol_OBS} shows the temporal variation of magnetic flux
and various types of energies during the whole time of the
simulation. Note that all the energies are normalized by the potential
field energy $E_{0}$ at $t = 0$. It shows a pre-eruption stage from
$t=0$ to around $32t_{s}$ and an eruption stage after $t=32t_{s}$,
with onset of the eruption indicated by a rapid rise of the kinetic
energy (\Fig~\ref{paraevol_OBS}d). The total unsigned magnetic fluxes
in both the observation and simulation (\Fig~\ref{paraevol_OBS}a)
decrease overall. As a result, both the total and potential magnetic
energies decay gradually (\Fig~\ref{paraevol_OBS}b). Nevertheless, the
free magnetic energy shows a continuous increase before the eruption
and experiences a fast decrease during the eruption, which is
consistent with expectation that free energy is stored before eruption
and released by eruption. The total released free magnetic energy
amounts to approximately $0.15E_0$ or $2\times 10^{32}$~erg if scaled
to the realistic value of the magnetic field, which is sufficient to
power a typical X-class
flare~\citep{emslieGLOBALENERGETICSTHIRTYEIGHT2012}. This evolution
pattern is clearer in the profile of the ratio of total magnetic
energy to the potential field energy, i.e., the degree of
non-potentiality. It first increases monotonically to a critical value
of $1.9$, and then decreases with the onset of the eruption. By
considering that the boundary driving is sped up by $68.6$ times, and
thus in the quasi-static driving phase, $t_s$ corresponds to 2 hours
in real time, the simulated eruption start time of $t=32t_s$
corresponds to $64$ hours. This agrees strikingly well with the onset
time of the observed flare, which is $t=65$~h. During the eruption,
the evolution speed is determined by the corona itself, so the
interval from the eruption onset time to the peak time of kinetic
energy is around $14t_s$ (24~min), and this matches closely the
observed duration from the GOES flare onset time to its peak time,
which is also 24~min (\Fig~\ref{paraevol_OBS}d).

\Fig~\ref{fig3_OBS} shows snapshots before the eruption; panels a and
b for the evolution of magnetic flux distribution at the bottom
surface in the simulation and {that from the observed magnetograms,
  respectively}, panel c for a synthetic images of coronal emission
from the coronal magnetic configuration, panel d for the corresponding
AIA~94~{\AA} images, panel e for current density normalized by
magnetic field strength (i.e., $J/B$) on a vertical cross section, and
panel f for $J/B$ on a horizontal cross section. To mimic the emission
of the coronal magnetic structure, we generated synthetic images of
coronal emission from current density using a method similar to that
proposed by~\citet{cheungMETHODDATADRIVENSIMULATIONS2012}. As it is
believed that the coronal loops generally reflect the structure of the
magnetic field lines rooted in the photosphere, we first traced a
sufficiently large number ($\sim 10^6$) of field lines with their
footpoints uniformly distributed at the bottom surface. All the field
lines are traced with fixed step of $720$~km. Then on each field line,
all the line segments are assigned with a proxy value of emission
intensity represented by the averaged square of current density along
this field line, by simply assuming that the Ohmic dissipation of the
currents heats the corona. Finally, the total emission along the line
of sight (here simply along the $z$ axis) is obtained by integrating
all the emission intensity along the $z$ axis, which forms the final
synthetic image.

The magnetic configuration at the initial time is a sheared arcade
core enveloped by an overlying, nearly current-free field. As the main
sunspot rotates counterclockwise, the coronal magnetic configuration
expands progressively and form a prominent reverse S shape
(\Fig~\ref{fig3_OBS}c). The synthetic coronal emission has a high
degree of resemblance with the AIA observations at the corresponding
times (comparing \Fig~\ref{fig3_OBS}c and d), in particular, the
central sigmoid structure immediately prior to the eruption. A clear
signature of current sheet formation can be seen in the evolution of
current density in cross sections of the volume (\Fig~\ref{fig3_OBS}e,
f and Supplementary Video 2). Note that the current density is
normalized by the magnetic field strength (i.e., $J/B$) to emphasize
thin layers with strong current. Initially the current density is
volumetric, and gradually a narrow layer with enhanced density
emerges, becomes progressively thinner. To characterize this evolution
of the current layer, we have measured its thickness, which is defined
at the location where it is thinnest. As can be seen in the variation
of the current layer thickness with time in \Fig~\ref{paraevol_OBS}d,
the thickness of the current layer decreases all the way until the
onset of the eruption. At the time of $t = 32t_s$, the thin current
layer extends from the bottom to a height of $5L_{s}$ with a thickness
of around $3\Delta$ (here $\Delta = 0.72$~Mm is the finest grid
resolution). This is the critical time point when the current sheet
reaches beyond the grid resolution and the built-in resistivity arises
to trigger fast reconnection in the current sheet, which initiates the
eruption.

\Fig~\ref{fig4_OBS} shows three snapshots of 3D magnetic field lines
and 2D sliced $J/B$ during the eruption (see also Supplementary Video
2). The eruption of the magnetic field creates a large-scale MFR
through the continuous magnetic reconnection in the current sheet. The
existence of such an MFR in this event has been confirmed by in-situ
observation in the interplanetary space of the CME from this
AR~\citep{kilpuaEstimatingMagneticStructure2021}. Ahead of the MFR,
the eruption drives a fast magnetosonic shock, which is shown by the
thin arc of the current density on the top of the MFR.

It should be noted that our data-driven simulation cannot reproduce
exactly the evolution of the photospheric magnetic field (in its three
components). {The difference between the simulated and observed
  magnetograms are clearly seen by comparing \Fig~\ref{fig3_OBS}a with
  b}. In \Fig~\ref{compare_mag_fig} we quantified the difference
between the simulated and the observed magnetic field $B_z$ at the
bottom surface by computing the correlation coefficient (CC) between
them. Both the two simulations (V2D driven by only the horizontal
flow, and V3D as driven by the full vector velocity) show the CC
decreases with time. The CC on average is around $0.8$ for the V3D
simulation, but it is lower ($0.7$) for the V2D simulation due to the
omission of the vertical component of the photospheric velocity. In
\Fig~\ref{compare_mag_fig}b, we have also compared the cumulative
magnetic energy as computed from the bottom boundary ($\vec v$ and
$\vec B$) in the numerical model with that derived directly from the
observed data ($\vec v$ and the observed $\vec B$). The V3D simulation
can reproduce quite well (with a relative error of around 10\%) the
cumulative energy as derived directly from the observed $\vec B$.

\section{Conclusions and discussion}\label{con}
In this paper, based on data-driven MHD simulations, we have studied
the process of a rotating sunspot in AR NOAA~12158 causing a major
solar eruption. The simulation is started from an MHD equilibrium
constructed for a time of 65~h before the eruption, and then is driven
by the photospheric flow field that is recovered from a time sequence
of SDO/HMI vector magnetograms. The whole evolution in the simulation
is in good agreement with observations, especially in the timing of
transition from the pre-eruption phase to the eruption onset, the
duration of the eruption, and the formation of a prominent sigmoid
before the eruption. The simulation shows that, before the eruption,
the degree of nonpotentiality of the coronal magnetic field, as
measured by the ratio of the total magnetic energy to the
corresponding potential field energy, is increased monotonically by
the surface rotation flow, while the kinetic energy keeps a small
value, as the MHD system evolves quasi-statically. At a critical time,
there is a clear transition from the quasi-static state to an eruptive
phase in which the kinetic energy impulsively rises and the free
magnetic energy releases quickly. Our simulations demonstrated that
through the successive rotation of the sunspot, the coronal field is
sheared with a vertical current sheet created progressively, and the
eruption is initiated by fast reconnection at the current sheet, which
producing a highly twisted flux rope during the eruption, forming a
CME. This supports the fundamental mechanism of solar eruption
initiation as recently established based on an idealized bipolar
magnetic configuration~\citep{jiangFundamentalMechanismSolar2021,
  bianHomologousCoronalMass2022,
  bianNumericalSimulationFundamental2022}.

There is a long standing controversy on the pre-eruption magnetic
structure and the initiation mechanism. Some argue that a magnetic
flux rope should exist before eruption and its ideal instability alone
can initiate eruption~\citep{kliemTorusInstability2006,
  torokConfinedEjectiveEruptions2005, fanOnsetCoronalMass2007,
  aulanierFORMATIONTORUSUNSTABLEFLUX2010,
  kliemCATASTROPHEINSTABILITYERUPTION2014}, while others argue that
there is no need of flux rope since reconnection of sheared arcade can
also initiate eruption~\citep{mooreOnsetMagneticExplosion2001,
  antiochosModelSolarCoronal1999, wyperUniversalModelSolar2017,
  jiangFundamentalMechanismSolar2021}. The results of our data-driven
simulation for the studied AR support the latter one. Our findings for
the pre-eruption magnetic configuration also confirm many previous
studies of the same event. For example, some nonlinear force-free
field extrapolations from the vector magnetograms suggested that the
pre-eruption field is a sheared arcade rather than a well-defined
pre-existing flux rope~\citep{vemareddySUNSPOTROTATIONDRIVER2016,
  duanComparisonTwoCoronal2017,
  heQuantitativeCharacterizationMagnetic2022}. Moreover, a magnetic
flux rope insertion method was employed
by~\citet{shenPrecursorPhaseXclass2022} and they also did not
construct a flux rope before the flare, but found two J-shaped sheared
arcades and overlying arcade in agreement with the AIA
observations. Recently, \citet{gouCompleteReplacementMagnetic2023}
made a comprehensive analysis of the motions of the complex flare
ribbons of this flare, and also argued that the flare reconnection
builds up totally the erupting flux rope, which agrees with our
simulation. Although some lowing-lying filaments were observed at the
flare site~\citep{dudikSLIPPINGMAGNETICRECONNECTION2016} which may
indicate a pre-existing flux rope, they were almost undisturbed during
the flare and thus played no role in the eruption initiation.

Last we note that our data-driven simulation does not reproduce
exactly the evolution of the photospheric magnetic field (in its three
components) because of the inherent limitation of the DAVE4VM code as
well as the smoothing of the data that is required for the limited
grid resolutions. In the DAVE4VM code the derived velocity (along with
the magnetic field) only approximately satisfies the vertical
component of the magnetic induction equation, and moreover the
equation is used in a statistical way rather than a point-by-point
sense~\citep{schuckTrackingVectorMagnetograms2008}.  It is hopeful
that more accurate data-driven modelling with higher resolution can be
realized using advanced photospheric velocity recovering method with
more constraints from observation, for example, the PDFI electric
field inversion approach that solves for all three components of the
induction equation precisely and takes Doppler velocity into
account~\citep{fisherPDFISSElectric2020,hoeksemaCoronalGlobalEvolutionary2020a,fisherCoronalGlobalEvolutionary2015}.




\acknowledgments
  This work is jointly supported by Shenzhen Science and Technology
  Program (Grant No. RCJC20210609104422048), Shenzhen Technology
  Project JCYJ20190806142609035, Shenzhen Key Laboratory Launching
  Project (No. ZDSYS20210702140800001), Guangdong Basic and Applied
  Basic Research Foundation (2023B1515040021) and National Natural
  Science Foundation of China (NSFC 42174200). Data from observations
  are courtesy of NASA SDO/AIA and the HMI science teams. The
  computational work was carried out on TianHe-1(A), National
  Supercomputer Center in Tianjin,
  China. 



\end{document}